\documentclass[twocolumn,prb,aps,floatfix,showpacs,superscriptaddress]{revtex4}
\usepackage{graphicx}
\usepackage{dcolumn}
\usepackage{bm}
\usepackage{color}

\begin{document}

\title{Atomically thin binary V-V compound semiconductor: a first-principles study}
\author{Weiyang Yu}
\email[e-mail address:] {yuweiyang@hpu.edu.cn}
\affiliation{School of Physics and Chemistry, Henan Polytechnic University, Jiaozuo 454000, China}
\affiliation{International Laboratory for Quantum Functional Materials of Henan, Zhengzhou University, Zhengzhou, 450001, China}
\author{Zhili Zhu}
\affiliation{International Laboratory for Quantum Functional Materials of Henan, Zhengzhou University, Zhengzhou, 450001, China}
\author{Chun-Yao Niu}
\affiliation{International Laboratory for Quantum Functional Materials of Henan, Zhengzhou University, Zhengzhou, 450001, China}
\author{Xiaolin Cai}
\affiliation{School of Physics and Chemistry, Henan Polytechnic University, Jiaozuo 454000, China}
\affiliation{International Laboratory for Quantum Functional Materials of Henan, Zhengzhou University, Zhengzhou, 450001, China}
\author{Wei-Bing Zhang}
\email[e-mail address:] {zhangwb@csust.edu.cn}
\affiliation{School of Physics and Electronic Sciences, Changsha University of Science and Technology, Changsha, 410004, China}
\date{\today}
\begin{abstract}

\textbf{Abstract}
Searching the novel 2D semiconductor is crucial to develop the next-generation low-dimensional electronic device. Using first-principles calculations, we propose a class of unexplored binary V-V compound semiconductor (PN, AsN, SbN, AsP, SbP and SbAs) with  monolayer black phosphorene ($\alpha$)  and blue phosphorene ($\beta$) structure. Our phonon spectra and room-temperature molecular dynamics (MD) calculations indicate that all compounds are very stable. Moreover, most of compounds are found to present a moderate energy gap in the visible frequency range, which can be tuned gradually by in-plane strain. Especially,  $\alpha$-phase V-V compounds have a direct  gap  while $\beta$-SbN, AsN, SbP, and SbAs may be promising  candidates of 2D solar cell materials due to a wide gap separating acoustic and optical phonon modes. Furthermore, vertical heterostructures can be also built using lattice matched $\alpha$($\beta$)-SbN and phosphorene, and both vdW heterostructures are found to have intriguing direct band gap. The present investigation not only  broads  the scope of layered group V semiconductors but also  provides an unprecedented route for the potential applications of 2D V-V families in optoelectronic and nanoelectronic semiconductor devices.

\end{abstract}


\maketitle

\section{\label{intruduction}Intruduction}
Two-dimensional (2D) semiconductors of group V elements including phosphorene, arsenene, and antimonene have been rapidly attracting interest on account of their significant wide-range fundamental band gap,  high and anisotropic carrier mobility, linear
dichroism, and anisotropic thermal conductivities.\cite{Li,Liu,Guan,Kamal,Kou,Wang2,Zhang2,Zhang3,YY,Y,TH} The outstanding properties make these systems as very favorable contenders for 2D electronics applications beyond graphene and transition metal dichalcogenides (TMDs).\cite{Elias,Ganesh,Radisavljevic}  The success of group V monolayer motivates the ongoing search for related 2D materials with unusual properties.  Recently, phosphorene has been extended notably by introducing isoelectronic IV-VI compounds,\cite{Arunima,Zhu2} it is thus intriguing to see whether the binary V-V compound monolayers can be achieved.

As established in case of phosphorus, group V elements and their compounds can form interesting  2D layered structures such as black-phosphorene-like [$\alpha$-phase, (No.64)] and blue-phosphorene-like [$\beta$-phase, (No.166)], in which the atomic layers are binding with  vdW interaction. It's well known that graphene and phosphorene can be mechanically exfoliated from  graphite and bulk black phosphorus.\cite{Reich} It is thus viable that the layered black-phosphorene-like ($\alpha$-phase) and blue-phosphorene-like ($\beta$-phase) of AsP and SbAs structures can be made into monolayer AsP and SbAs. \cite{Krebs} Very recently, Kou \emph{et al.} have investigated monolayer arsenic and its compound SbAs, which can be seen as a single-layer of  bulk compound with space group of R3m. Except SbAs, P, As and Sb can also be pairwise combined to form various binary compounds. It is expected that these binary V-V materials will exhibit unforeseen properties that present invaluable opportunities for innovative applications.

In this work, we perform a systematical first-principles study of the as yet unexplored monolayer binary group V compounds, such as PN, AsN, SbN, SbP, AsP, and SbAs both in  $\alpha$- and $\beta$-allotropes. The detailed equilibrium geometry and electronic structure are obtained and compared.  Additionally, the stabilities of these monolayer compounds are also analyzed by energetics, vibrational spectra and room-temperature molecular dynamics (MD) simulations. Furthermore, taking AsP as an example, we also show that the electronic structure of these binary compounds can be tuned effectively by the in-layer strain. Finally, a vertical heterostructure stacked with  phosphorene and lattice-matched SbN  (SbN/P) is built to illustrate the possible device application.

\section{\label{comp}Computational details}

Our DFT calculations have been performed using Vienna \emph{ab initio} simulation package (VASP) code.\cite{Kresse2} We used the Perdew-Burke-Ernzerhof (PBE)\cite{Perdew} exchange-correlation functional for the GGA. The projector augmented wave (PAW) method\cite{Kresse3} was employed to describe the electron-ion interaction. A kinetic energy cutoff of the plane-wave basis set was used to be 500 eV.  All the structures are optimized until Hellmann-Feynman residual forces less than 0.01 eV/{\AA}. In the MD calculations, (5$\times$5) supercells are employed to minimize the constraint of periodic boundary condition and the temperature was kept at 300K for 6 ps with a time step of 2 fs in the moles-volume-temperature (NVT) ensemble.  Tetrahedron method was used with a quick projection scheme in the calculations of the density of state (DOS). For the calculations of the band structures, we used Gaussian smearing in combination with a small width of 0.05 eV, and the path of integration in first Brillouin zone is along Y(0.0, 0.5, 0.0)$\rightarrow$$\Gamma$(0.0, 0.0, 0.0)$\rightarrow$X(0.5, 0.0, 0.0)$\rightarrow$M(0.5, 0.5, 0.0) for $\alpha$-phase, and along $\Gamma$(0.0, 0.0, 0.0)$\rightarrow$M(0.0, 0.5, 0.0)$\rightarrow$K(1/3, 2/3, 0.0)$\rightarrow$$\Gamma$(0.0, 0.0, 0.0) for $\beta$-phase. We used an adequate number of $k$-points for compounds with the different supercell sizes, equivalent to 9$\times$9$\times$1 Monkhorst-Pack \cite{Monkhosrt} sampling for (1$\times$1) unit cell . In order to avoid spurious interactions between periodic images of the layer, a vacuum spacing perpendicular to the plane was employed to be larger than $\sim$15 {\AA}.

\section{\label{res}Results and discussion}
Six different binary compounds in $\alpha$- and $\beta$ structures shown in Fig. 1 (a) and (b) are considered in the present paper.  The monolayer structures have been optimized using DFT with the PBE exchange-correlation functional, as discussed in the methods section. The calculated structural parameters and cohesive energy of monolayer were listed in Table 1. As shown in the table,  we can see that the 2D lattice of $\alpha$-phase is spanned by the orthogonal Bravais lattice parameters $a_{1}$=4.22, 4.20, 4.42, 4.60, 4.18, 4.30 {\AA} and $a_{2}$=2.08, 2.98, 3.30, 3.60, 4.08, 4.20 {\AA}, respectively. While the 2D hexagonal lattice of $\beta$-phase is spanned by two Bravais lattice parameters $a_{1}$=$a_{2}$=2.78, 3.03 3.33, 3.47, 3.86, 4.00 {\AA}. The corresponding  parameters of black phosphorene  and  blue phosphorene are ($a_{1}$=4.59{\AA}, $a_{2}$=3.31{\AA}),\cite{Yu1}  and  ($a_{1}$=$a_{2}$=3.33{\AA}).\cite{Yu2} The cohesive energies of $\alpha$-phase are -6.76, -6.02, -5.73, -4.31, -4.51, -4.26 eV/atom, respectively. And those of $\beta$-phase are -6.23, -5.58, -5.31, -4.58, -4.33, -4.92 eV/atom, respectively.  This suggested the most stable allotrope for all Nitrogen-contained compound is $\alpha$-phase. Whereas for phosphide,  $\beta$-AsP seems to be more stable by about 268 meV/atom than $\alpha$-AsP, which is different from counterpart structures of the phosphorene monolayer, in which  the energy of  $\alpha$ structures is lower. AsP trends to adopt the $\beta$ structure.

\begin{figure}[htb]
\includegraphics[width=8cm]{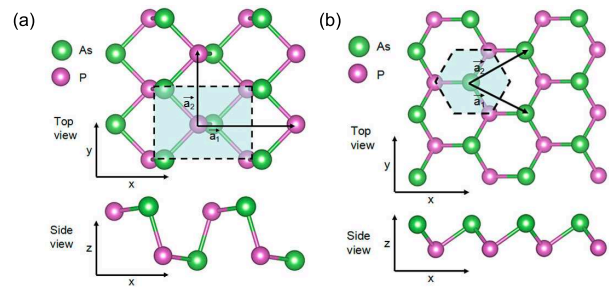}
\caption{(Color online) Geometric structures of (a) $\alpha$-AsP and (b) $\beta$-AsP monolayers of ball-and-stick models. And the Wigner-Seitz cell indicated by the shaded region.}
\end{figure}

\begin{table}
\caption{Structural parameters of 12 kinds of 2D V-V semiconductors. $a_{1}$ and $a_{2}$ are the lattice parameters as defined in Fig. 1. $E_{coh}$  is the cohesive energy defined as $E_{coh}$=$E_{tot}$-$E_{X}$-$E_{Y}$, where $E_{tot}$ is the total energy of the monolayer compounds, and $E_{X}$ and $E_{Y}$ are the energy of the isolated X and Y atoms, respectively.}
\begin{tabular}{p{1.4cm}p{1.4cm}p{2cm}p{1.4cm}p{1.6cm}}
\hline
\hline
Phase          & Name & $a_{1}$ (\AA) & $a_{2}$ (\AA) &$E_{coh}$ (eV/atom)  \\
\hline
               & PN      & 4.22        & 2.08         &  -6.76   \\
               & AsN     & 4.20        & 2.98         &  -6.02  \\
$\alpha$-phase & SbN     & 4.42        & 3.30         &  -5.73  \\
               & AsP     & 4.60        & 3.60         &  -4.31  \\
               & SbP     & 4.18        & 4.08         &  -4.51   \\
               & SbAs    & 4.30        & 4.20         &  -4.26  \\
\hline
               & PN      & $a_{1}$=$a_{2}$=2.78  &    &  -6.23  \\
               & AsN     & $a_{1}$=$a_{2}$=3.03  &    &  -5.58  \\
$\beta$-phase  & SbN     & $a_{1}$=$a_{2}$=3.33  &    &  -5.31  \\
               & AsP     & $a_{1}$=$a_{2}$=3.47  &    &  -4.58   \\
               & SbP     & $a_{1}$=$a_{2}$=3.86  &    &  -4.33   \\
               & SbAs    & $a_{1}$=$a_{2}$=4.00  &    &  -4.92  \\
\hline
\hline
\end{tabular}
\end{table}

\begin{figure*}[htb]
\includegraphics[width=15cm]{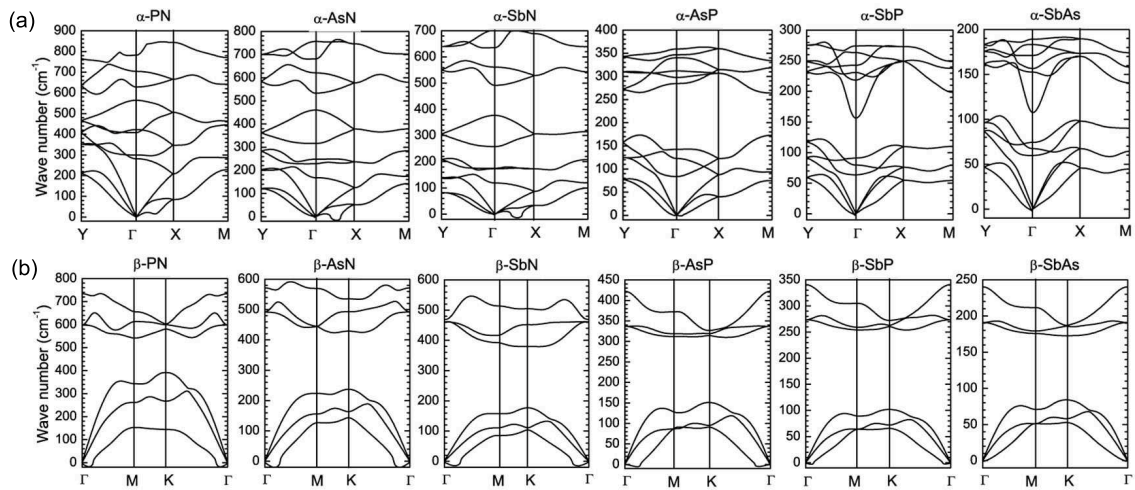}
\caption{Vibrational phonon spectra of (a) $\alpha$-phase and (b) $\beta$-phase of PN, AsN, SbN, AsP, and SbAs monolayers, respectively.}
\end{figure*}

\begin{figure*}[htb]
\includegraphics[width=15cm]{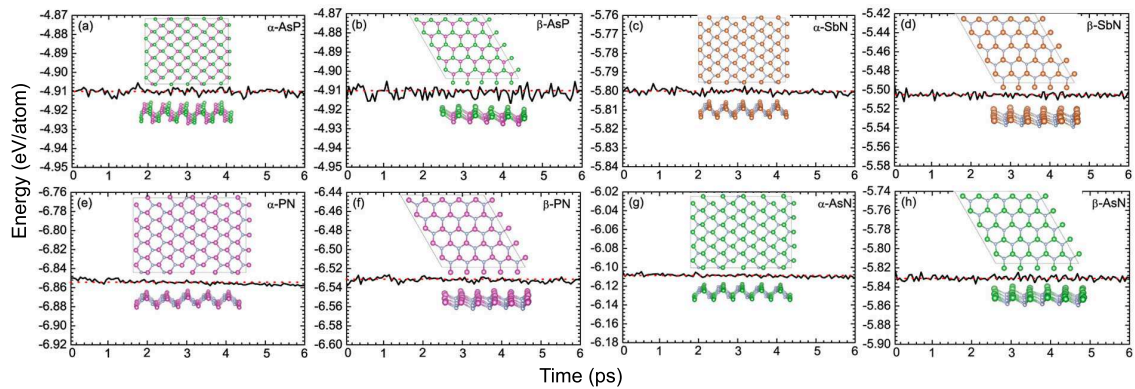}
\caption{(color on line) Relationships of total energy and time during room-temperature MD simulations of  (a) $\alpha$-AsP, (b) $\beta$-AsP, (c) $\alpha$-SbN, (d) $\beta$-SbN, (e) $\alpha$-PN, (f) $\beta$-PN, (g) $\alpha$-AsN, and (h) $\beta$-AsN, respectively. The final geometric structures at the end of 6 ps are also shown. The MD simulations are performed at T=300K last for 6 ps.}
\end{figure*}

 To further confirm the stability of these monolayer binary compounds, we also perform vibrational phonon spectra calculations. The calculated vibration spectra of $\alpha$- and $\beta$- PN, AsN, SbN, AsP, SbP, and SbAs monolayers are presented in Fig. 2. We find there are basically no imaginary frequencies in whole Brillouin zone for both $\alpha$- and $\beta$- phases, which means these compounds are dynamically stable and can exist as free-standing 2D crystals. The ``U'' shape features in $\alpha$- AsN, $\alpha$- SbN, $\beta$- PN, $\beta$-AsN, $\beta$-SbN, and $\beta$-AsP spectra near the $\Gamma$-point is a signature of the flexural acoustic mode but not instability. This is usually hard to converge in 2D layers, which is also found in other similar system.\cite{Shengbai} Furthermore, the phonon spectra of compounds in the same structure are rather similar, reflecting a very similar bonding character.  It is interesting to notice that the acoustic and optical modes are well separated in $\beta$-phase compounds, indicating good optic properties. In detail, the $\beta$-phase phonon spectra exhibit a wide gap ($\omega_{g}$) separating acoustic and optical modes of $\sim$150, 180, 220, 170, 150, and 90 cm$^{-1}$ for $\beta$- PN, AsN, SbN, AsP, SbP, and SbAs, respectively. During the real solar cell application, photon-excited electron-hole pairs  will loose most of their energy in the form of heat by getting scattered off the lattice, and emitting optical phonon. These optical modes then decay into acoustic modes, which diffuse faster and carry the heat away. As a result, most high energy photo-electrons relax to the conduction band edge before being extracted. As shown in the figure, the phonon spectra gaps of  $\beta$- SbN, AsP, SbP and SbAs ($\sim$220, 170, 150, 90 cm$^{-1}$, respectively) are significantly greater than the hardest acoustic mode ($\sim$180, 150, 100, 80 cm$^{-1}$, respectively). This can avoid such a ``Klemens decay" effectively,\cite{Konig} which suggests that these materials can be used in high-efficient solar cells application.

\begin{figure*}[htb]
\includegraphics[width=15cm]{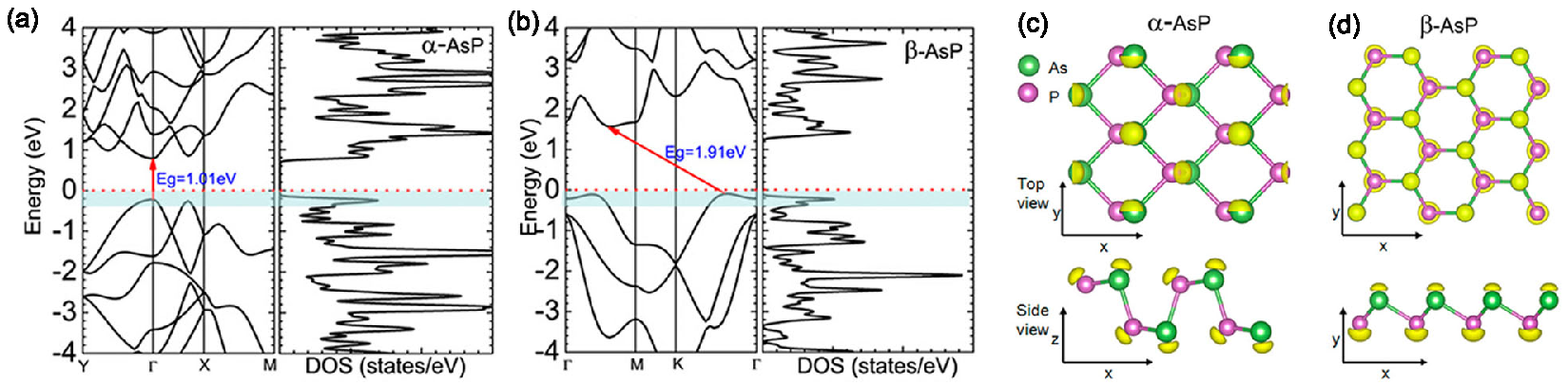}
\caption{(color on line) (a) and (b) The band structures and the density of states (DOS) of $\alpha$ and $\beta$-AsP. The energy range between $E_{f}$-0.2 eV  and $E_{f}$, indicated by the green shading, is used to identify valence frontier states. The Fermi level is set at zero. (c) and (d) Band decomposed charge density $\rho_{vb}$ associated with states in the shading energy range. }
\end{figure*}

\begin{figure*}[htb]
\includegraphics[width=15cm]{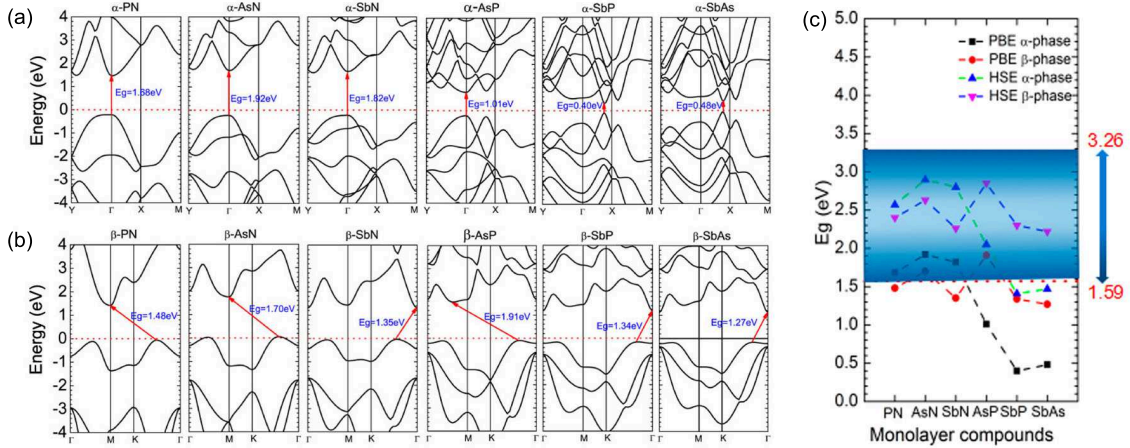}
\caption{(Color online) Band structures of (a) $\alpha$-phase and (b) $\beta$-phase for different monolayer compounds, along with (c) the values of band gap with PBE and HSE calculations. The Fermi level is set at zero.}
\end{figure*}

\begin{figure*}[htb]
\includegraphics[width=15cm]{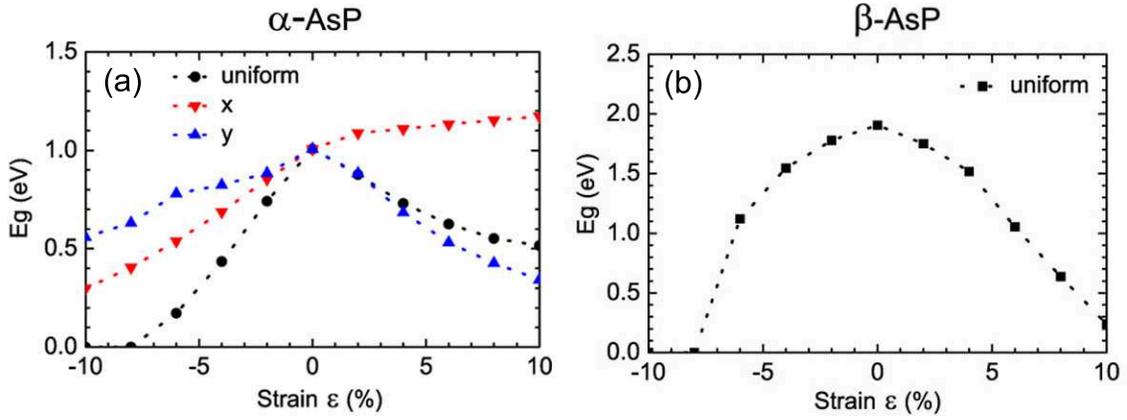}
\caption{ Electronic band gaps of (a) $\alpha$-AsP and (b) $\beta$-AsP monolayers as a function of the in-layer strain, from -10\% to 10\%, with an interval scale of 2\%. The dot line are guides to the eye.}
\end{figure*}

\begin{figure*}[htb]
\includegraphics[width=15cm]{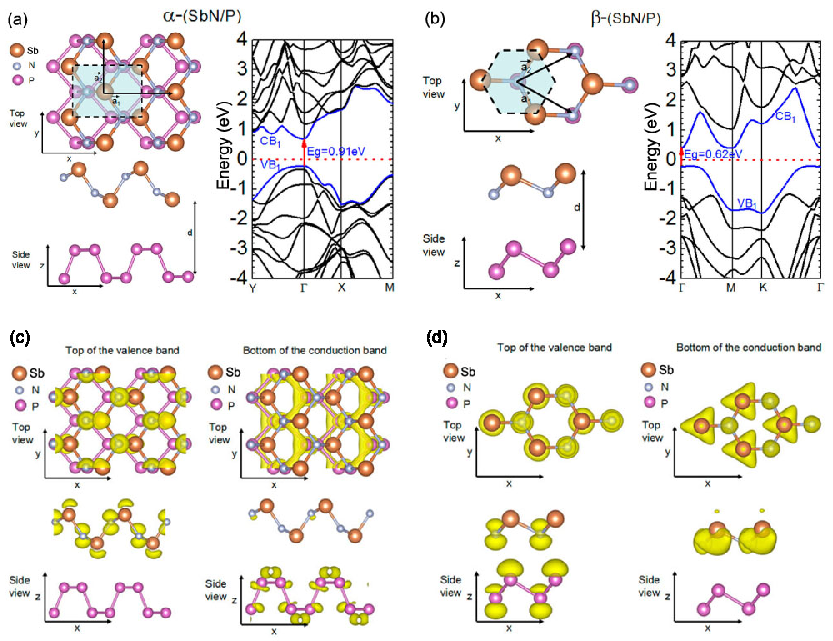}
\caption{ (Color online) Optimum geometric and electronic band structures of (a) $\alpha$-(SbN/P) bilayer and (b) $\beta$-(SbN/P) bilayer, respectively. The optimum stacking of the SbN and the phosphorene monolayer in the $\alpha$-(SbN/P) bilayer in (a) is AB and that in the $\beta$-(SbN/P) bilayer in (b) is AA. Band decomposed charge densities of bilayer $\alpha$-(SbN/P) (c) and $\beta$-(SbN/P) (d), respectively. $\rho_{vb}$=0.08 e/\AA$^3$ contours are superposed with ball-and-stick models of related structures.}
\end{figure*}

Furthermore, we have also performed \emph{ab initio} molecular dynamics (MD) simulations to examine the thermal stability of the single-layer compounds at room temperature. The evolution of geometric structures and potential energies of $\alpha$- and $\beta$- AsP, SbN, PN, and AsN during simulation  at room temperature are given in Fig. 3. As seen in the figure, the total energies of the systems oscillate just around the energy less than 5 meV/atom, and the structures kept almost the same as their equilibrium structures, indicating that these monolayers  are dynamic stable indeed at room temperature. This also suggests that the monolayer binary V-V compound can be realized experimentally even at room temperature.

Now turn our attention to the electronic structure of monolayer binary V-V compounds. The calculated band structure, along with the corresponding density of states for $\alpha$-AsP, as presented in Fig. 4 (a), suggest that the fundamental band gap value $E_{g}$=1.01 eV is slightly larger than that of  $\alpha$-P counterpart (0.91eV)\cite{Yu1}. The band structure near the top of the valence band shows a remarkable anisotropy comparing the $\Gamma$$\rightarrow$X and $\Gamma$$\rightarrow$Y directions. The DFT-based fundamental band gap $E_{g}$=1.91 eV of monolayer $\beta$-AsP is also larger than that of blue phosphorene counterpart (1.82eV)\cite{Yu2}. The band structure in the symmetric honeycomb lattice of $\beta$-AsP is rather isotropic, as seen in Fig. 4 (b). The top of the valence band is flat, leading to a heavy hole mass and a density of states (DOS) peak in that region. As we know, the character of frontier states is not only of interest for a microscopic understanding of the conduction channels but also of great concern for the design of optimal contacts.\cite{Tom¨¢nek}  The charge density corresponds to the frontier states near the top of the valence band in Fig. 4 (c) and (d), which associates with the energy ranging from E$_f$-0.2 eV  to  E$_f$ highlighted by the green shading in the band structure of $\alpha$-AsP in Fig. 4 (a) and that of $\beta$-AsP in Fig. 4 (b). The valence frontier states of $\alpha$-AsP in Fig. 4 (c) and $\beta$-AsP in Fig. 4 (d) are similar and a typical  lone pair electron state are found in these frontier states, which is similar to those of phosphorene. \cite{Rudenko}

The calculated band structures of $\alpha$- and $\beta$- PN, AsN, SbN, AsP, SbP, and SbAs are shown in Fig. 5 (a) and (b), respectively. Generally, the band structures of $\alpha$-phase indicate direct band gap, while that of $\beta$-phase displays indirect band gap. In detail, the direct band gaps of $\alpha$- PN, AsN, SbN, AsP, SbP, and SbAs are 1.68, 1.92, 1.82, 1.01, 0.40, and 0.48 eV respectively. While that of $\beta$- PN, AsN, SbN, AsP, SbP, and SbAs are 1.48, 1.70, 1.35, 1.91, 1.34, and 1.27, respectively. The unique direct/indirect gap features of $\alpha$-phases and  $\beta$-phases  have been evidenced using elementary group theory\cite{Li2,Ribeiro-Soares} to originate from the special puckered and buckled honeycomb structures of $\alpha$-phase and $\beta$-phase, respectively. Since the fundamental band gap is known to be underestimated significantly  in the semi-local PBE method,  we also use hybrid HSE06 functional  to obtain accurate electronic structures. The band gaps with PBE as well as HSE calculations of $\alpha$- and $\beta$- PN, AsN, SbN, AsP, SbP, and SbAs are shown in Fig. 5 (c). As seen in Fig. 5 (c), most band gaps of the monolayer compounds lie in the visible light region (1.59-3.26 eV), implying possible optoelectronic applications.

In analogy to black and blue phosphorenes, the fundamental band gap values of $\alpha$- and $\beta$- phases also depend sensitively on the in-layer strain. On account of their nonplanarity, accordion-like in-layer stretching or compression of AsP structure may be achieved at little energy cost, as shown in the Supporting Informations (Fig. S1). The energy cost is particularly low for a deformation along the soft x-direction, requiring $\sim$60 meV/atom to induce a 10\% in-layer strain. We believe that in view of the soft structure, similar strain values may be achieved in the course of epitaxial growth on particular disproportionate matrixes. We also note that such tensile strain values have been achieved in suspended graphene membranes experimentally that are much more flexible to stretch.\cite{Lee,Huang}  We can believe that strain engineering is a feasible way to effectively tune the fundamental band gaps of these systems.

The band gaps of $\alpha$- and $\beta$- as a function of in-layer strain are shown in Fig. 6. For $\alpha$-AsP, as seen in Fig. 6 (a), the band gap decreases(increases) slightly when the structure is compressed(stretched) along x direction, while the band gap always decreases both in compression and in stretching along y direction. When uniform  strain are applied, the band gap will also decreases. The band gap will reduce to 1.0 eV with a 8\% compression. As for $\beta$-AsP,  as seen in Fig. 6 (b),  the fundamental band gap decreases during both stretching and compression. Within the $\pm$8\% range, we find that the band gap may be tuned in the range from 0 to 1.9 eV. This high degree of band gap tunability in AsP appears particularly charming for potential applications in flexible electronics. For $\alpha$-AsP, when compressing from 6\% to 10\%, the material gradually turned  to metal. The detailed band-strain relationships of $\alpha$- and $\beta$- AsP monolayers were shown in Supporting Informations (Fig. S2).

The P-N junction is one of the fundamental building blocks for modern electronics. With recent discoveries of atomically thin materials, layer-by-layer stacking (vertically stacked) or lateral interfacing (in-plane interconnected) heterojunction has been reported, \cite{Huang2,Gong,Tian,Padilha,Hu,Cai,Nathaniel} which indicates the traditional semiconductor devices can be scaled down to atomic thicknesses.

It is intriguing that the geometry and lattice constants of group V-V compounds and phosphorene are very similar. It is thus possible that these V-V compounds  could interface naturally with  phosphorene in lateral and vertical heterojunctions, thus further promoting the tunability of their electronic properties. In Fig. 7, we present geometrical and electronic structuress for bilayers consisting of SbN and phosphorene (termed with SbN/P) in both $\alpha$- and $\beta$- phases with lattice mismatch less than 5\% as the simplest examples of vertical heterojunctions. We have optimized the bilayer structures assuming commensurability, i.e., setting the primitive unit cells of each monolayer to be the same. The optimum geometries of these  $\alpha$-(SbN/P) bilayers are shown in Fig. 7. We find the interlayer interaction in the two bilayer systems seem to be rather weak, amounting to 30 meV/atom based on our DFT-PBE calculations.  It is interesting to notice that both vdW heterojunctions have a distinguish direct gap although the monolayer $\beta$-SbN presents a indirect gap. The $\alpha$-(SbN/P) bilayer is direct band gap semiconductors of 0.91 eV, almost equal to that of black phosphorene. Whereas the band dap of $\beta$-(SbN/P) bilayer is also direct band gap of 0.62 eV, which is smaller than that of either isolated monolayer.We further analyze the frontier states of both systems.  As shown in Fig. 7 (c), the  states of  $\alpha$-(SbN/P) near VBM and CBM are dominated by SbN and  phosphorene, respectively. For $\beta$-(SbN/P), we find a substantial rehybridization  between the adjacent SbN and phosphorene layers near VBM while that in the bottom of conduction region is from SbN dominantly, as seen in Fig. 7 (d). This interesting hybridization will lead to the reduction of gap in $\beta$-(SbN/P) systems. Since both SbN and phosphorene are rather flexible, they may also form in-layer heterostructures at little or no energy penalty. We also constructed four types of SbNP$_{2}$ heterostructures and  a reduced energy gaps were found in all heterostructures. The detailed geometry and electronic structure are given in Supporting Informations (Fig. S3).

\section{\label{con}Conclusions}
 Using first-principles DFT calculations, we have identified yet unrealized structural phases of  PN, AsN, SbN, AsP, SbP, and AsSb, in the black-phosphorus-like $\alpha$ phase and the almost equally stable blue-phosphorus-like $\beta$ phase. We have verified that all the studied monolayer compounds have a good energetic and dynamic stability. Particularly, the calculations of phonon spectra indicate that $\beta$- SbN, AsN, SbP, and SbAs may be underlying candidates of 2D solar cell materials. The band structure calculations show that $\alpha$-phase compounds display a direct band gap, while $\beta$-phases compounds display a significant indirect band gap. Both $\alpha$-phase and $\beta$-phase compounds depend on the in-layer strain strongly. This strain dependence of bandgap offers an unprecedented tunability in structural and electronic properties of group V compounds. Furthermore, we also construct lateral and vertical heterostructures using phosphorene and  SbN . Both vertical $\alpha$-(SbN/P) and $\beta$-(SbN/P) heterojunctions are found to have direct energy gap ,  which may be used to design novel 2D heterojuction devices. The present investigation suggests that binary Group V compounds consists of a large family of layered semiconductor compounds with an unprecedented richness in structural and electronic properties.

\begin{acknowledgments}
This work was supported by the National Natural Science Foundation of China (Grant Nos. 11504332, 11304288, and 11404171).
\end{acknowledgments}

\emph{Supporting Information:} The energy-strain and band-strain relationships of $\alpha$- and $\beta$- AsP monolayers, along with the calculated geometric and electronic structures of in-plane heterostructures are available free.

\end{document}